\documentclass[final,a4paper]{aastex702}

\usepackage{amsmath}
\usepackage{amssymb}
\usepackage{bm}
\usepackage{booktabs}

\shorttitle{Morphological Identification of MHD Regimes}
\shortauthors{Zhao et al.}

\begin{document}

\title{Morphological Identification of Dynamical Regimes in MHD Turbulence with ScaleAware-JEPA}

\correspondingauthor{Guang-Xing Li and Keping Qiu}

\author[0000-0003-0596-6608]{Mengke Zhao}
\affiliation{School of Astronomy and Space Science, Nanjing University, 163 Xianlin Avenue, Nanjing 210023, Jiangsu, People’s Republic of China}
\affiliation{Key Laboratory of Modern Astronomy and Astrophysics (Nanjing University), Ministry of Education, Nanjing 210023, Jiangsu, People’s Republic of China}
\email{mkzhao@nju.edu.cn}

\author[0000-0003-3144-1952]{Guang-Xing Li}
\affiliation{South-Western Institute for Astronomy Research, Yunnan University, Kunming 650091, People’s Republic of China}
\email[show]{gxli@ynu.edu.cn}

\author[0000-0002-5093-5088]{Keping Qiu}
\affiliation{School of Astronomy and Space Science, Nanjing University, 163 Xianlin Avenue, Nanjing 210023, Jiangsu, People’s Republic of China}
\affiliation{Key Laboratory of Modern Astronomy and Astrophysics (Nanjing University), Ministry of Education, Nanjing 210023, Jiangsu, People’s Republic of China}
\email[show]{kpqiu@nju.edu.cn}

\begin{abstract}
The morphology of turbulent interstellar gas reflects the coupled action of turbulence, magnetic fields, and self-gravity, but density amplitude alone does not uniquely specify dynamical state. We use ScaleAware-JEPA to learn multiscale structural coordinates from density alone and test whether these learned coordinates organize dynamical states more clearly than density-based conditioning. Clump-like, filament-like, and diffuse-like latent neighborhoods show an ordered progression in turbulent velocity scaling that is not reproduced by density-selected populations. Velocity and magnetic fields are withheld during training and used afterward as post-training physical diagnostics: MHD states occupy ordered but overlapping regions of the learned representation, and dimensionless dynamical balances vary coherently across the full latent atlas. These results show that a representation learned from density morphology can acquire a geometry that is systematically organized by established MHD diagnostics. Multiscale morphology therefore provides a useful coordinate for dynamical state beyond density amplitude alone.
\end{abstract}


\vspace{-4mm}

\begin{figure}[!b]
\centering
\includegraphics[width=\textwidth]{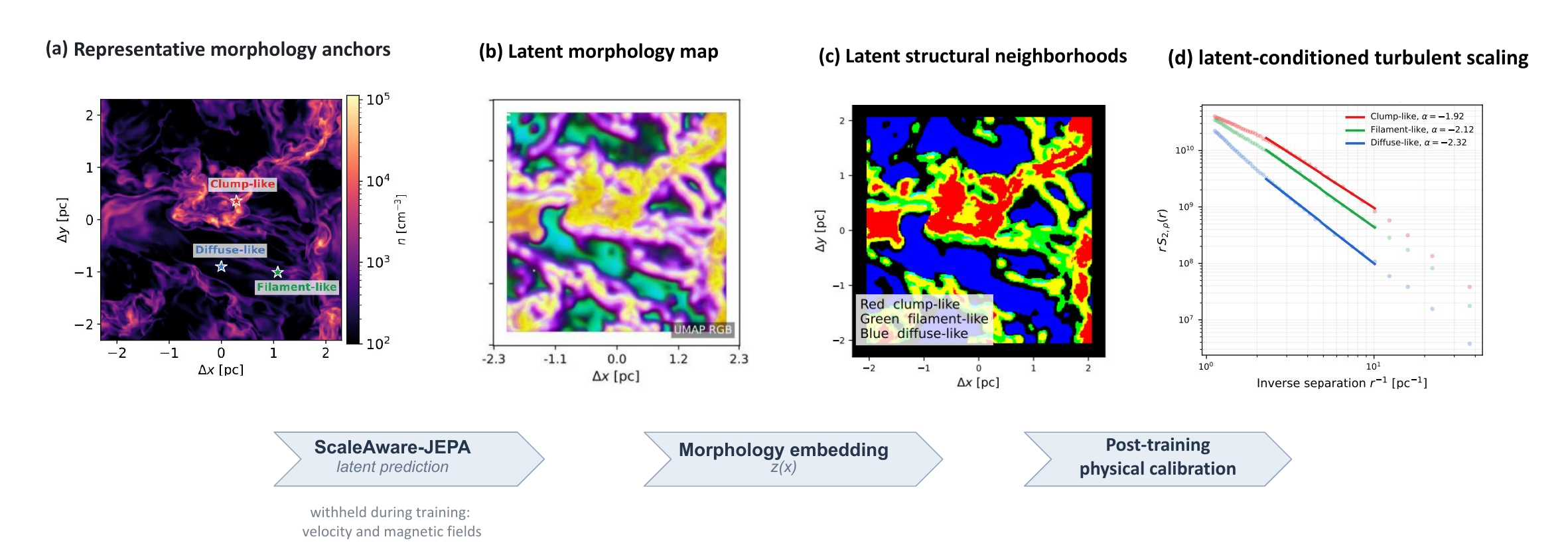}
\caption{From density morphology to dynamical differentiation.
The workflow arrows summarize the analysis: ScaleAware-JEPA learns a
multiscale morphological representation from density alone, while velocity
and magnetic fields are withheld during training and used afterward for
physical calibration.
(a) Spatial back-mapping of three representative latent anchors to
clump-like, filament-like, and diffuse-like density morphologies.
(b) Continuous latent morphology map obtained from the UMAP-RGB
representation of the learned coordinates; UMAP is used here only to
visualize the latent organization.
(c) Latent structural neighborhoods associated with the three anchors,
shown by their upper-tertile similarity back-maps; mixed colors indicate
overlapping neighborhoods.
(d) Density-weighted second-order turbulent scaling within the same latent
neighborhoods. The fitted slopes steepen monotonically from the clump-like
to filament-like to diffuse-like environments. The corresponding
density-conditioned comparison is presented in Figure~\ref{fig:spectra}.}
\label{fig:framework}
\end{figure}

\section{Introduction}\label{sec:intro}

The star-forming interstellar medium (ISM) is a hierarchical MHD system whose dynamics reflect the scale-dependent competition among supersonic turbulence, magnetic fields, and self-gravity \citep{1981MNRAS.194..809L,1987ARA&A..25...23S,2004ARA&A..42..211E,2004RvMP...76..125M,2007ARA&A..45..565M,2012A&ARv..20...55H}. The classical linewidth--size and density--size relations already pointed to correlated motions and structures across scale \citep{1981MNRAS.194..809L}, while more recent MHD studies show that magnetic, kinetic, and gravitational terms can exchange dynamical importance continuously with density and scale \citep{2024ApJ...976..209Z,2025MNRAS.542.3246L,2026arXiv260906900Z}. Much of the physical information in such a system is therefore encoded in relations rather than in isolated amplitudes \citep{zhao2026gravitydrivenemergencemultifractaldensity}: how turbulent fluctuations vary with separation, how magnetic field strength changes with density, and how competing dynamical terms compare.

Here we ask whether a coordinate learned from multiscale density morphology retains physical information absent from the scalar density \citep{zhao2026scalevectoralignmentscaleawareframework}. Unlike an analysis built directly from prescribed physical variables, a learned representation has no a priori physical interpretation. Compressible MHD turbulence offers a useful test: similar densities occur in structurally and dynamically different environments, while velocity and magnetic fields can be excluded from representation learning and used afterward as post-training diagnostics. Density morphology is itself shaped by the coupled action of turbulence, magnetic fields, and gravity, so a physically informative density-learned representation should organize the established velocity and magnetic relations. Dimensionless parameters are particularly useful for this purpose because they measure the relative importance of competing processes; for example, $M_A\sim(E_k/E_B)^{1/2}$ traces the continuous transition between magnetically and kinetically dominated gas \citep{2024ApJ...976..209Z,2026arXiv260906900Z}.

For this test we use ScaleAware-JEPA, which learns a dense representation from density alone while velocity and magnetic fields remain withheld. Joint-Embedding Predictive Architectures (JEPA; \citealt{2022LeCunAMI}) learn predictive representations by estimating latent representations of unobserved targets from observed context rather than reconstructing targets directly in input space; I-JEPA demonstrated this principle for self-supervised image representation learning \citep{2023arXiv230108243A}. ScaleAware-JEPA adapts the same idea to continuous physical fields by tying the prediction geometry to an explicit hierarchy of physical scale \citep{2026arXiv260629723L}. Here that hierarchy is supplied by Constrained Diffusion Decomposition (CDD; \citealt{2022ApJS..259...59L}), described in Section~\ref{sec:sajepa}. The model receives no morphological labels, velocity field, or magnetic field during training.

We test the learned coordinates first with turbulent velocity scaling, then with post-training magnetic and Alfv\'enic diagnostics, and finally across the full latent atlas using dimensionless dynamical balances. Velocity and magnetic fields enter only after training and serve as post-training probes of the latent geometry. The key question is whether the density-learned representation separates dynamical regimes more clearly than density-based conditioning, and whether established MHD physics can in turn calibrate the structure it discovers.

\section{ScaleAware-JEPA: Structural Coordinates from Multiscale Density}\label{sec:sajepa}

ScaleAware-JEPA constructs a dense structural coordinate from the multiscale organization of a two-dimensional density field. We apply it to the self-gravitating, supersonic MHD simulation described in Appendix~\ref{app:simulation}; no velocity or magnetic information enters the representation learning.

For the reported ScaleAware-JEPA experiments, CDD organizes the density field into a five-level, pixel-registered scale pyramid at diffusion-scale coordinates $s=2,4,8,16,$ and $32$ pixels \citep{2022ApJS..259...59L,2026arXiv260629723L}. We denote the corresponding components by
\begin{equation}
 n(\bm{x}) \longrightarrow \{C_s(\bm{x})\}_{s=2,4,8,16,32}.
\end{equation}
The final CDD channel contains the coarsest component together with the unresolved residual. These same scale coordinates organize both the encoder input and the geometry of the latent-prediction task. The aligned CDD components are jointly encoded, while the context-mask footprint is tied to the diffusion scale of each component and the target patch remains fixed.

The model learns by predicting hidden target representations from visible multiscale context rather than reconstructing density pixels \citep{2022LeCunAMI,2023arXiv230108243A}. A context (online) encoder receives the visible multiscale field, and a predictor estimates the latent representation at hidden target positions. The corresponding target representation is produced from the unmasked field by a target encoder whose parameters follow the online encoder through an exponential-moving average. The training objective is
\begin{equation}
 {\cal L}=\lambda_{\rm pred}{\cal L}_{\rm pred}
 +\lambda_{\rm spread}{\cal L}_{\rm spread},
 \label{eq:sajepa_loss}
\end{equation}
where ${\cal L}_{\rm pred}$ is the JEPA latent-prediction term,
\begin{equation}
 {\cal L}_{\rm pred}\propto
 \left\|\widehat{\bm z}_{\rm target}-\bm z_{\rm target}\right\|_2^2,
 \qquad
 \widehat{\bm z}_{\rm target}=g\!\left(\bm z_{\rm context},M,s\right),
\end{equation}
and ${\cal L}_{\rm spread}$ is a weak spread regularizer that maintains channel-wise variation in the projected context representation and prevents representational collapse. For the reported MHD configuration, $\lambda_{\rm pred}=50$ and $\lambda_{\rm spread}=5$ \citep{2026arXiv260629723L}. Here $M$ denotes the context/target mask geometry and $s$ the CDD scale coordinate. Each valid spatial location is consequently associated with a learned descriptor of its multiscale density environment.

The resulting representation assigns a 32-dimensional latent vector to every valid spatial location,
\begin{equation}
 \bm{z}(\bm{x})\in\mathbb{R}^{32}.
\end{equation}
At inference, the trained ScaleAware-JEPA model produces a dense latent map over the full field. PCA and UMAP provide back-mappable views of the learned geometry: localized regions of the projected atlas can be traced to their original spatial coordinates and interpreted through the corresponding density morphology \citep{2026arXiv260629723L}. Figure~\ref{fig:framework} summarizes the physical logic of the analysis.
Representative positions selected in the latent atlas back-map to
clump-like, filament-like, and diffuse-like density morphologies
(panel a). The UMAP-RGB back-map provides a continuous visualization
of the learned morphological organization (panel b), while similarity
to the three representative anchors defines overlapping latent structural
neighborhoods across the physical domain (panel c). The same neighborhoods
show clearly different density-weighted turbulent scaling
(panel d), providing the first post-training dynamical calibration of
the learned coordinates. Velocity and magnetic fields are withheld
throughout representation learning and introduced only afterward as
dynamical diagnostics.

\section{Results and Discussion}\label{sec:results}

\subsection{Turbulent scaling distinguishes dynamical states across morphological environments}\label{subsec:spectra}

We begin with turbulent scaling: turbulence generates and continually reshapes the density structure. Since the Kolmogorov picture, scale-dependent velocity statistics have provided a basic description of turbulent transport and cascade \citep{1941DoSSR..30..301K,1995tlnk.book.....F}. Molecular-cloud turbulence is supersonic, strongly compressible, and magnetized, so shocks, compressive motions, and the forcing mixture affect both velocity scaling and the density field \citep{2007ApJ...665..416K,2010A&A...512A..81F}. If the density-learned structural coordinates carry dynamical information, environments separated in the learned representation should also show different scale-dependent velocity statistics.

Following the ScaleAware-JEPA latent-atlas workflow, we inspect the three-dimensional UMAP projection and select three representative localized positions in the learned manifold. Because the dense latent field is pixel registered, each selected position maps back to a unique spatial pixel; their back-mapped density environments reveal clump-like, filament-like, and diffuse-like morphology. The associated similarity back-maps define three latent neighborhoods, with each field thresholded independently at its 66.67th percentile. Their overlap therefore traces continuity between neighboring structural environments in the latent atlas.

A density-weighted second-order structure function provides the scale-dependent velocity diagnostic inside each latent environment. For a structural selection $M$,
\begin{equation}
 S_{2,\rho}(r\mid M)=
 \frac{\left\langle \sqrt{\rho_i\rho_j}\,|\bm v_i-\bm v_j|^2\right\rangle_{r,M}}
 {\left\langle \sqrt{\rho_i\rho_j}\right\rangle_{r,M}},
 \label{eq:s2}
\end{equation}
where both endpoints of a pair lie in the selected environment. We plot $rS_{2,\rho}(r)$ against inverse separation $r^{-1}$ and fit
\begin{equation}
 rS_{2,\rho}(r)\propto (r^{-1})^{\alpha},
 \qquad \alpha=-(\zeta_2+1)
 \quad\mathrm{for}\quad S_2(r)\propto r^{\zeta_2}.
 \label{eq:es2}
\end{equation}
The fit spans $5\le r/\Delta x\le25$, corresponding to $0.09$--$0.45$ pc for $\Delta x=0.018$ pc.

The latent environments exhibit a clear ordering in turbulent scaling. The best-fit slopes are
\begin{equation}
 \alpha_C=-1.92,\qquad \alpha_F=-2.12,\qquad \alpha_D=-2.32,
\end{equation}
for the clump-like, filament-like, and diffuse-like selections. The same snapshot, estimator, density weighting, and fit interval yield an almost uniform steepening from compact to diffuse latent environments. The shallower clump-like relation indicates relatively enhanced velocity increments toward smaller separations, whereas the diffuse-like environment shows the steepest scale dependence. Because velocity was withheld from SA-JEPA, this ordered sequence directly connects the density-learned latent coordinate to distinct turbulent states.

\begin{figure}
\centering
\includegraphics[width=\textwidth]{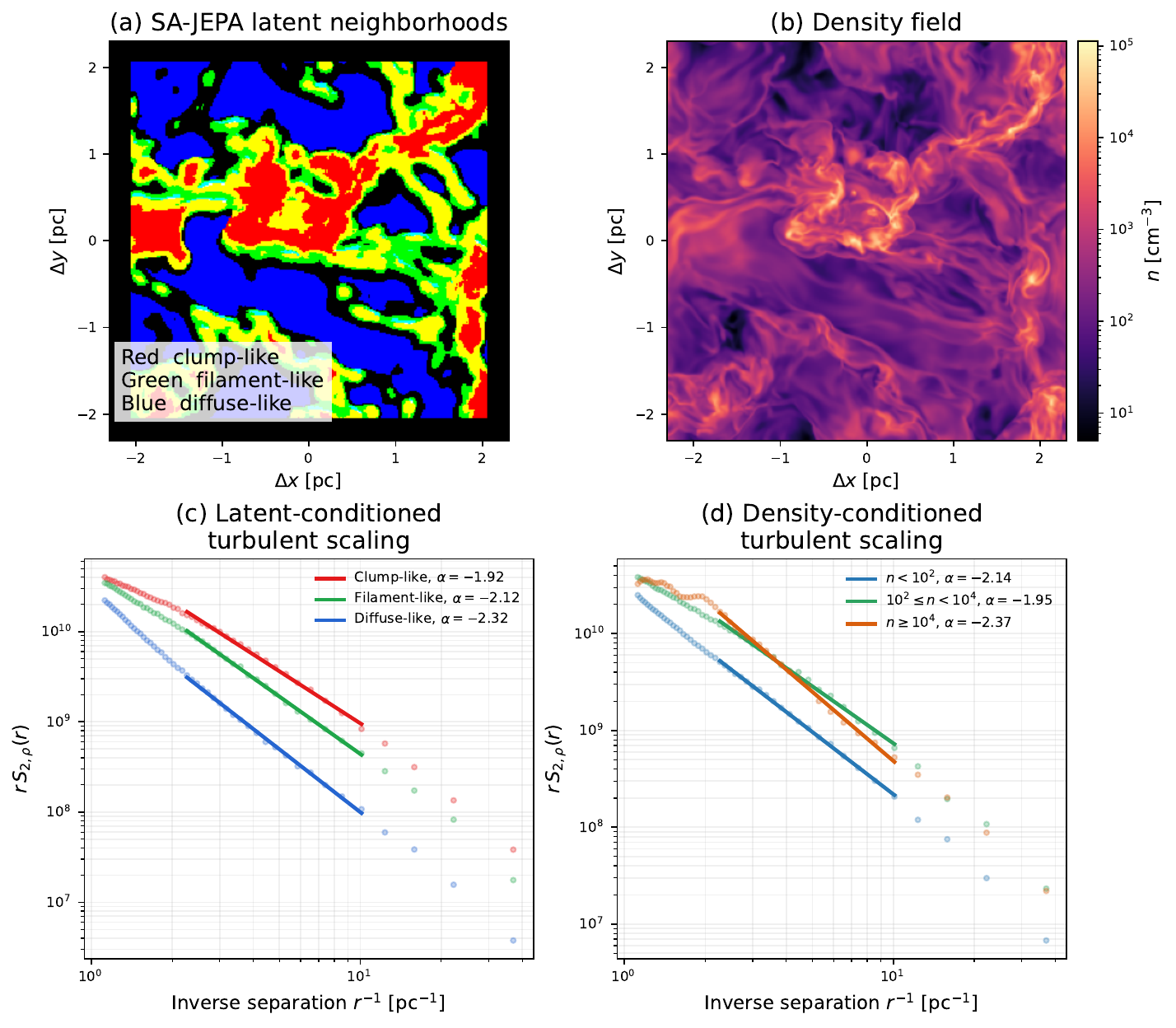}
\caption{Latent structural conditioning versus scalar density conditioning. (a) Non-exclusive upper-tertile similarity back-maps around three representative UMAP-selected latent anchors; mixed colors indicate overlapping latent neighborhoods. (b) Number-density field in the same slice. (c) Density-weighted second-order velocity scaling, displayed as $rS_{2,\rho}(r)$ against inverse separation $r^{-1}$. The fitted slopes steepen monotonically from $-1.92$ to $-2.12$ to $-2.32$. (d) The same estimator applied to three number-density bins gives $-2.14$, $-1.95$, and $-2.37$, a non-monotonic sequence.}
\label{fig:spectra}
\end{figure}

The density-conditioned comparison does not show the same ordering. The bins $n<10^2\,{\rm cm^{-3}}$, $10^2\le n<10^4\,{\rm cm^{-3}}$, and $n\ge10^4\,{\rm cm^{-3}}$ give slopes $-2.14$, $-1.95$, and $-2.37$ (Figure~\ref{fig:spectra}d), with the intermediate-density bin being the shallowest. Thus the monotonic clump--filament--diffuse sequence is not recovered by density bins, indicating that the multiscale morphological representation captures structural information beyond density amplitude alone.

\subsection{Magnetic coupling and the continuous Alfv\'enic transition}\label{subsec:mhdcal}

We next ask whether the same density-learned environments organize the magnetic field, a degree of freedom absent from hydrodynamic turbulence. Magnetic tension changes the cascade geometry and produces scale-dependent anisotropy relative to the local field \citep{1995ApJ...438..763G,2003MNRAS.345..325C}; observations of molecular clouds likewise show ordered fields and anisotropic turbulent motions when magnetic stresses are dynamically important \citep{2011MNRAS.411.2067L}. The $B$--$\rho$ relation records how the magnetic field responds to compression and, in the gradual-transition picture, how magnetic and kinetic importance change with density \citep{2024ApJ...976..209Z,2026arXiv260906900Z}.

The latent environments trace an ordered, continuous sequence in the $B$--$\rho$ plane. For this distributional comparison, we use exclusive assignments to avoid double counting pixels shared by neighboring latent environments: each pixel is assigned to the strongest of the three prototype similarities provided that similarity exceeds its 66.67th-percentile threshold. The full slice is shown in gray, while colored contours trace the 20\%, 50\%, and 80\% highest-probability-density regions of the conditional KDEs. Diffuse-like gas occupies lower $\rho$ and $B$, filament-like gas lies at intermediate values, and clump-like gas extends toward higher density and stronger field. The overlapping contours and progressively displaced KDE cores connect these environments through a continuous magnetic sequence.

Removing the leading density dependence exposes additional organization in the coupled magnetic--turbulent state. We construct median relations in 24 bins of $\log\rho$ spanning the 1st--99th percentiles and define
\begin{align}
 \Delta\log B &= \log B-\log B_{\rm med}(\rho),\\
 \Delta\log\sigma_v &= \log\sigma_v-\log\sigma_{v,\rm med}(\rho).
\end{align}
After detrending, the residual distributions remain overlapping yet displaced from one another. Using equal-size samples, the pooled-covariance Mahalanobis distances are $D_M=0.21$, $0.85$, and $0.62$ for C--F, F--D, and C--D, while the corresponding two-dimensional overlap coefficients are 0.75, 0.60, and 0.52. The largest residual displacement occurs between filament-like and diffuse-like environments. The latent sequence therefore reflects joint magnetic--turbulent organization in addition to the leading density trend.

The Alfv\'en Mach number provides a dimensionless measure of the same magnetic--turbulent balance. Within an $8\times8$-pixel aperture in the central slice, we use density-weighted velocity moments,
\begin{equation}
 \bar\rho=\langle\rho\rangle,\qquad
 \bar v_q=\frac{\langle\rho v_q\rangle}{\bar\rho},\qquad
 \sigma_v^2=\sum_{q=x,y,z}\left(\frac{\langle\rho v_q^2\rangle}{\bar\rho}-\bar v_q^2\right),
\end{equation}
and define
\begin{equation}
 M_A=\frac{\sigma_v}{v_A}
 =\sqrt{4\pi\bar\rho}\,\frac{\sigma_v}{|\overline{\bm B}|}.
 \label{eq:ma2d}
\end{equation}
Because $M_A^2$ measures kinetic relative to magnetic energy under the adopted normalization, $M_A$ directly characterizes their dynamical balance. The resulting Alfv\'enic distributions overlap, while their main peaks and medians shift in the same direction: $M_A=0.36$ for diffuse-like gas, $0.75$ for filament-like gas, and $1.22$ for clump-like gas. The sequence spans strongly sub-Alfv\'enic diffuse-like gas, filament-like gas near the Alfv\'enic transition, and mildly super-Alfv\'enic clump-like gas. Together with the substantial overlap, this ordering is consistent with a gradual magnetic-to-kinetic transition instead of a separation into discrete MHD phases \citep{2024ApJ...976..209Z,2025MNRAS.542.3246L,2026arXiv260906900Z}.

\begin{figure}
\centering
\includegraphics[width=\textwidth]{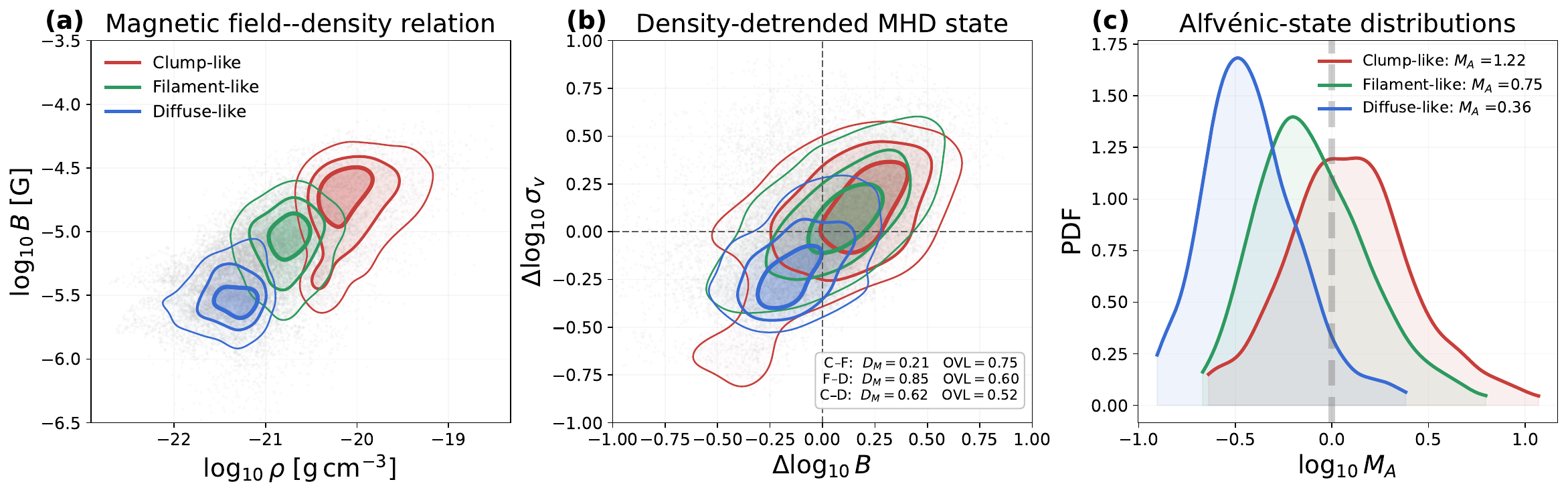}
\caption{Post-training magnetic and Alfv\'enic calibration of the latent environments. (a) Conditional populations in the $B$--$\rho$ plane. Colored contours trace 20\%, 50\%, and 80\% highest-probability-density regions of the KDEs for the exclusive clump-like, filament-like, and diffuse-like assignments. Their overlap and ordered displacement indicate a continuous sequence rather than three disconnected states. (b) The same environments after removal of the global density trends, in the plane $(\Delta\log B,\Delta\log\sigma_v)$, using density-weighted local velocity dispersion. The inset reports pooled-covariance Mahalanobis distances $D_M$ and two-dimensional overlap coefficients OVL from equal-size samples. (c) Slice-local Alfv\'en Mach-number distributions obtained from the same $8\times8$-pixel apertures. Median values shift from $0.36$ (diffuse-like) through $0.75$ (filament-like) to $1.22$ (clump-like).}
\label{fig:mhdcal}
\end{figure}

\subsection{Dynamical balance across the full latent atlas}\label{subsec:manifold}

The three prototype environments sample only part of a continuous representation. We therefore extend the comparison across the full 32-dimensional latent space and use the three-dimensional UMAP projection as a common display coordinate \citep{2018arXiv180203426M}; the clump-like, filament-like, and diffuse-like anchors remain reference points within this atlas. This atlas view checks whether the learned morphology also organizes the competition among turbulence, magnetic fields, and gravity beyond the three selected neighborhoods.

Turbulence, magnetic fields, and gravity define three pairwise dynamical balances. We characterize them with dimensionless measures of kinetic--magnetic, turbulent--gravitational, and gravity--magnetic competition. Independent diagnostics of the full three-dimensional density cube place the characteristic structural scale near $6$--$6.4$ pixels: the density power spectrum shows a local spectral maximum near $6$ pixels, while the CDD-based mass-weighted effective-scale distribution peaks at $6.4$ pixels (Appendix~\ref{app:estimators}; Figure~\ref{fig:scale_diag}). The SA-JEPA CDD pyramid uses $s=2,4,8,16,$ and $32$ pixels, so we adopt the nearest scale, $s_*=8$ pixels, corresponding to $\ell_*=8\Delta x=0.144$ pc. This choice is consistent with the characteristic scale used by \citet{2024ApJ...976..209Z}. We use the Alfv\'en Mach number to characterize kinetic--magnetic balance,
\begin{equation}
 M_A=\sqrt{4\pi\bar\rho}\,\frac{\sigma_{v,3D}}{|\overline{\bm B}|}
 \sim\left(\frac{E_k}{E_B}\right)^{1/2},
 \label{eq:ma3d}
\end{equation}
a fixed-scale turbulent virial proxy for turbulent--gravitational balance \citep{1992ApJ...395..140B},
\begin{equation}
 \alpha_{\rm vir}=\frac{5\sigma_{1D}^2R_{\rm eff}}{GM},
 \label{eq:avir}
\end{equation}
and the normalized mass-to-flux proxy
\begin{equation}
 \lambda=2\pi\sqrt{G}\,\frac{\Sigma}{|\overline{\bm B}|}
 \label{eq:lambda}
\end{equation}
measures gravitational loading relative to magnetic flux in the standard criticality normalization \citep{1976ApJ...210..326M,1978PASJ...30..671N}. The local geometry used for $M$, $R_{\rm eff}$, and $\Sigma$ is given in Appendix~\ref{app:estimators}.

Across the common unsmoothed population, the median Alfv\'en Mach number is $M_A=0.741$ with a 16--84 percentile range $0.384$--$1.69$; $64.8\%$ of locations are sub-Alfv\'enic and $35.2\%$ are super-Alfv\'enic. The turbulent--gravitational proxy has median $\alpha_{\rm vir}=46.5$ with a 16--84 percentile range $12.8$--$189$, placing most locations in a turbulence-dominated balance at this fixed $0.144$-pc scale. The mass-to-flux proxy has median $\lambda=0.159$ with a 16--84 percentile range $0.0669$--$0.469$; $94.2\%$ of locations are magnetically subcritical, while $5.8\%$ reach or exceed $\lambda=1$.

The physical fields are then summarized over local neighborhoods in the original 32-dimensional SA-JEPA representation. Their UMAP projections form coherent but non-identical regions of kinetic--magnetic, turbulent--gravitational, and gravity--magnetic balance. UMAP is used only as the common display coordinate; the global medians and percentile ranges quoted above are computed from the unsmoothed physical fields.

\begin{figure}
\centering
\includegraphics[width=\textwidth]{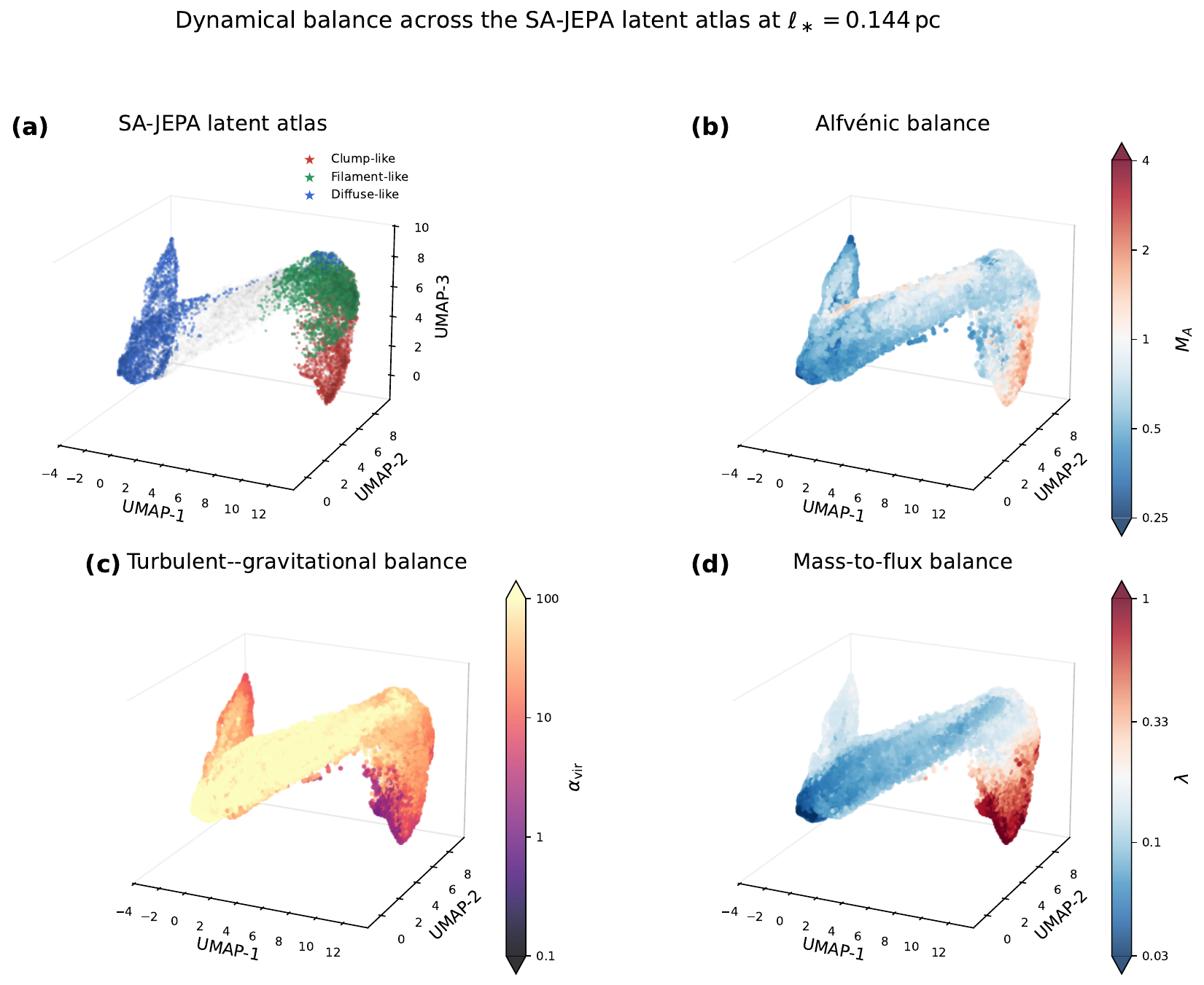}
\caption{Dimensionless MHD balance across the latent atlas at $\ell_*=8$ pixels $=0.144$ pc, the nearest SA-JEPA scale to the independently inferred $6$--$6.4$-pixel characteristic density scale (Figure~\ref{fig:scale_diag}). (a) Three-dimensional UMAP display of the full latent atlas. Colored overlays indicate representative clump-like, filament-like, and diffuse-like latent neighborhoods, while stars mark the UMAP-selected anchors; the overlapping regions sample a continuous latent representation. (b) Kinetic--magnetic balance traced by $M_A$. (c) Turbulent--gravitational balance traced by $\alpha_{\rm vir}$. (d) Gravity--magnetic balance traced by the mass-to-flux proxy $\lambda$. Colors in panels (b)--(d) show the median physical value among 96 nearest neighbors in the original 32-dimensional latent representation; UMAP provides the common display coordinates.}
\label{fig:manifold}
\end{figure}

\subsection{Physical calibration of learned morphological coordinates}\label{subsec:physicscoords}

A learned latent coordinate is not itself a conventional physical observable; its physical relevance has to be established from how withheld dynamical information is organized across it. In the present experiment, velocity, magnetic field, and dynamical-balance quantities never enter representation learning. They nevertheless follow consistent trends across the density-learned geometry. Representative neighborhoods show distinct turbulent scaling and occupy ordered, overlapping MHD states, while the full atlas contains coherent regions of kinetic--magnetic, turbulent--gravitational, and gravity--magnetic balance. The density-bin comparison shows that this ordering is not set by the density amplitude.

The learned geometry can therefore be calibrated with physical information that was withheld during training. We summarize this correspondence as
\begin{equation}
 \bm z\longrightarrow
 \left\{F(\ell\mid\bm z),\;P(\bm y\mid\bm z),\;\Pi(\bm z)\right\},
\end{equation}
where $F$ denotes a scale-dependent physical statistic, $P$ a conditional relation among fields, and $\Pi$ a set of dimensionless dynamical balances. Established MHD diagnostics thus provide a post-training test of the physical content of the learned coordinates: physically distinct regimes occupy overlapping, connected regions of a representation constructed from density morphology alone.

\section{Conclusions}\label{sec:conclusions}

Morphology-selected regions exhibit distinct dynamical properties. Turbulent scaling steepens monotonically from clump-like to filament-like to diffuse-like regions, with slopes of $-1.92$, $-2.12$, and $-2.32$, whereas density-selected populations do not reproduce this ordering. The same morphological sequence is reflected in the magnetic state: the median Alfv\'en Mach number increases from $0.36$ in diffuse-like regions through $0.75$ in filament-like regions to $1.22$ in clump-like regions, tracing a continuous magnetic-to-kinetic transition. After removal of the leading density dependence, the latent environments remain systematically displaced in joint magnetic--turbulent state. Across the full 32-dimensional latent representation, kinetic--magnetic, turbulent--gravitational, and gravity--magnetic balances form coherent structures at the adopted 8-pixel scale, the nearest SA-JEPA scale to the independently inferred $6$--$6.4$-pixel density structure.

These results show that multiscale density morphology carries dynamical information that is not recovered from density amplitude alone. Because velocity, magnetic field, and dynamical-balance quantities are withheld from representation learning, their organization in latent space provides a post-training physical calibration of the learned coordinates. ScaleAware-JEPA therefore supplies a morphology-based coordinate system whose physical content can be tested directly against established MHD diagnostics.







\bibliographystyle{aasjournalv7.1}
\bibliography{references}

@ARTICLE{2026arXiv260629723L,
       author = {{Li}, Guang-Xing},
        title = "{ScaleAware-JEPA: Latent Representation for Discovery in Multiscale Physical Fields}",
      journal = {arXiv e-prints},
         year = 2026,
        month = jun,
          eid = {arXiv:2606.29723},
        pages = {arXiv:2606.29723},
          doi = {10.48550/arXiv.2606.29723},
archivePrefix = {arXiv},
       eprint = {2606.29723},
 primaryClass = {cs.LG}
}

@ARTICLE{2026arXiv260906900Z,
       author = {{Zhao}, Mengke and {Li}, Guang-Xing and {Qiu}, Keping},
        title = "{A Unified Magnetohydrodynamic Scaling Relation for the Multiphase Interstellar Medium}",
      journal = {arXiv e-prints},
         year = 2026,
        month = sep,
          eid = {arXiv:2609.06900},
        pages = {arXiv:2609.06900},
          doi = {10.48550/arXiv.2609.06900},
archivePrefix = {arXiv},
       eprint = {2609.06900},
 primaryClass = {astro-ph.GA}
}

@ARTICLE{2024ApJ...976..209Z,
       author = {{Zhao}, Mengke and {Li}, Guang-Xing and {Qiu}, Keping},
        title = "{Slope of Magnetic Field--Density Relation as an Indicator of Magnetic Dominance}",
      journal = {\apj},
         year = 2024,
        month = dec,
       volume = {976},
       number = {2},
          eid = {209},
        pages = {209},
          doi = {10.3847/1538-4357/ad8b4d},
archivePrefix = {arXiv},
       eprint = {2409.02786},
 primaryClass = {astro-ph.GA}
}

@ARTICLE{2025MNRAS.542.3246L,
       author = {{Li}, Guang-Xing and {Zhao}, Mengke},
        title = "{Magnetic, kinetic, and transition regime: spatially segregated structure of compressive MHD turbulence}",
      journal = {\mnras},
         year = 2025,
        month = oct,
       volume = {542},
       number = {4},
        pages = {3246--3252},
          doi = {10.1093/mnras/staf1320},
archivePrefix = {arXiv},
       eprint = {2409.02769},
 primaryClass = {astro-ph.GA}
}

@ARTICLE{2022ApJS..259...59L,
       author = {{Li}, Guang-Xing},
        title = "{Multiscale Decomposition of Astronomical Maps: A Constrained Diffusion Method}",
      journal = {\apjs},
         year = 2022,
        month = apr,
       volume = {259},
       number = {2},
          eid = {59},
        pages = {59},
          doi = {10.3847/1538-4365/ac4bc4},
archivePrefix = {arXiv},
       eprint = {2201.05484},
 primaryClass = {astro-ph.IM}
}

@ARTICLE{2020ApJ...905...14B,
       author = {{Burkhart}, Blakesley and {Appel}, Sarah M. and {Bialy}, Shmuel and {Cho}, Jungyeon and {Christensen}, Anders J. and {Collins}, David and {Federrath}, Christoph and {Fielding}, Drummond B. and {Finkbeiner}, Douglas and {Hill}, Alex S. and {Ib{\'a}{\~n}ez-Mej{\'\i}a}, Juan C. and {Krumholz}, Mark R. and {Lazarian}, Alex and {Li}, Miao and {Mocz}, Philip and {Mac Low}, Mordecai-Mark and {Naiman}, Jillian and {Portillo}, Stephen K.~N. and {Shane}, Brian and {Slepian}, Zachary and {Yuan}, Ye},
        title = "{The Catalogue for Astrophysical Turbulence Simulations (CATS)}",
      journal = {\apj},
         year = 2020,
        month = dec,
       volume = {905},
       number = {1},
          eid = {14},
        pages = {14},
          doi = {10.3847/1538-4357/abc484},
archivePrefix = {arXiv},
       eprint = {2010.11227}
}

@ARTICLE{2015ApJ...808...48B,
       author = {{Burkhart}, Blakesley and {Collins}, David C. and {Lazarian}, Alex},
        title = "{Observational Diagnostics of Self-gravitating MHD Turbulence in Giant Molecular Clouds}",
      journal = {\apj},
         year = 2015,
        month = jul,
       volume = {808},
       number = {1},
          eid = {48},
        pages = {48},
          doi = {10.1088/0004-637X/808/1/48}
}

@ARTICLE{2012ApJ...750...13C,
       author = {{Collins}, David C. and {Kritsuk}, Alexei G. and {Padoan}, Paolo and {Li}, Hui and {Xu}, Hao and {Ustyugov}, Sergey D. and {Norman}, Michael L.},
        title = "{The Two States of Star-forming Clouds}",
      journal = {\apj},
         year = 2012,
        month = may,
       volume = {750},
       number = {1},
          eid = {13},
        pages = {13},
          doi = {10.1088/0004-637X/750/1/13},
archivePrefix = {arXiv},
       eprint = {1202.2594}
}

@ARTICLE{2010ApJS..186..308C,
       author = {{Collins}, David C. and {Xu}, Hao and {Norman}, Michael L. and {Li}, Hui and {Li}, Shengtai},
        title = "{Cosmological Adaptive Mesh Refinement Magnetohydrodynamics with Enzo}",
      journal = {\apjs},
         year = 2010,
        month = feb,
       volume = {186},
       number = {2},
        pages = {308--333},
          doi = {10.1088/0067-0049/186/2/308}
}

@ARTICLE{2012A&ARv..20...55H,
       author = {{Hennebelle}, Patrick and {Falgarone}, Edith},
        title = "{Turbulent molecular clouds}",
      journal = {\aapr},
         year = 2012,
        month = nov,
       volume = {20},
          eid = {55},
        pages = {55},
          doi = {10.1007/s00159-012-0055-y},
archivePrefix = {arXiv},
       eprint = {1211.0637}
}

@ARTICLE{2007ARA&A..45..565M,
       author = {{McKee}, Christopher F. and {Ostriker}, Eve C.},
        title = "{Theory of Star Formation}",
      journal = {\araa},
         year = 2007,
        month = sep,
       volume = {45},
        pages = {565--687},
          doi = {10.1146/annurev.astro.45.051806.110602},
archivePrefix = {arXiv},
       eprint = {0707.3514}
}

@ARTICLE{2004RvMP...76..125M,
       author = {{Mac Low}, Mordecai-Mark and {Klessen}, Ralf S.},
        title = "{Control of star formation by supersonic turbulence}",
      journal = {Reviews of Modern Physics},
         year = 2004,
        month = jan,
       volume = {76},
        pages = {125--194},
          doi = {10.1103/RevModPhys.76.125},
archivePrefix = {arXiv},
       eprint = {astro-ph/0301093}
}

@ARTICLE{1995ApJ...438..763G,
       author = {{Goldreich}, Peter and {Sridhar}, Sridhar},
        title = "{Toward a Theory of Interstellar Turbulence. II. Strong Alfv\'enic Turbulence}",
      journal = {\apj},
         year = 1995,
        month = jan,
       volume = {438},
        pages = {763--775},
          doi = {10.1086/175121}
}

@ARTICLE{2007ApJ...665..416K,
       author = {{Kritsuk}, Alexei G. and {Norman}, Michael L. and {Padoan}, Paolo and {Wagner}, Rick},
        title = "{The Statistics of Supersonic Isothermal Turbulence}",
      journal = {\apj},
         year = 2007,
        month = aug,
       volume = {665},
        pages = {416--431},
          doi = {10.1086/519443},
archivePrefix = {arXiv},
       eprint = {0704.3851}
}

@ARTICLE{2010A&A...512A..81F,
       author = {{Federrath}, Christoph and {Roman-Duval}, Julia and {Klessen}, Ralf S. and {Schmidt}, Wolfram and {Mac Low}, Mordecai-Mark},
        title = "{Comparing the statistics of interstellar turbulence in simulations and observations. Solenoidal versus compressive turbulence forcing}",
      journal = {\aap},
         year = 2010,
        month = mar,
       volume = {512},
          eid = {A81},
        pages = {A81},
          doi = {10.1051/0004-6361/200912437},
archivePrefix = {arXiv},
       eprint = {0905.1060}
}

@ARTICLE{1992ApJ...395..140B,
       author = {{Bertoldi}, Frank and {McKee}, Christopher F.},
        title = "{Pressure-confined Clumps in Magnetized Molecular Clouds}",
      journal = {\apj},
         year = 1992,
        month = aug,
       volume = {395},
        pages = {140--157},
          doi = {10.1086/171638}
}

@ARTICLE{1976ApJ...210..326M,
       author = {{Mouschovias}, Telemachos Ch. and {Spitzer}, Lyman Jr.},
        title = "{Note on the Collapse of Magnetic Interstellar Clouds}",
      journal = {\apj},
         year = 1976,
        month = dec,
       volume = {210},
        pages = {326--327},
          doi = {10.1086/154835}
}

@ARTICLE{1978PASJ...30..671N,
       author = {{Nakano}, Takenori and {Nakamura}, Takashi},
        title = "{Gravitational Instability of Magnetized Gaseous Disks}",
      journal = {\pasj},
         year = 1978,
        month = dec,
       volume = {30},
        pages = {671--679},
          doi = {10.1093/pasj/30.4.671}
}

@ARTICLE{1941DoSSR..30..301K,
       author = {{Kolmogorov}, A. N.},
        title = "{The Local Structure of Turbulence in Incompressible Viscous Fluid for Very Large Reynolds Numbers}",
      journal = {Doklady Akademii Nauk SSSR},
         year = 1941,
       volume = {30},
        pages = {301--305}
}

@BOOK{1995tlnk.book.....F,
       author = {{Frisch}, Uriel},
        title = "{Turbulence: The Legacy of A. N. Kolmogorov}",
         year = 1995,
    publisher = {Cambridge University Press},
      address = {Cambridge},
      doi = {10.1017/CBO9781139170666}
}

@ARTICLE{2023arXiv230108243A,
       author = {{Assran}, Mahmoud and {Duval}, Quentin and {Misra}, Ishan and {Bojanowski}, Piotr and {Vincent}, Pascal and {Rabbat}, Michael and {LeCun}, Yann and {Ballas}, Nicolas},
        title = "{Self-Supervised Learning from Images with a Joint-Embedding Predictive Architecture}",
      journal = {arXiv e-prints},
         year = 2023,
        month = jan,
          eid = {arXiv:2301.08243},
        pages = {arXiv:2301.08243},
archivePrefix = {arXiv},
       eprint = {2301.08243},
 primaryClass = {cs.CV}
}

@ARTICLE{2018arXiv180203426M,
       author = {{McInnes}, Leland and {Healy}, John and {Melville}, James},
        title = "{UMAP: Uniform Manifold Approximation and Projection for Dimension Reduction}",
      journal = {arXiv e-prints},
         year = 2018,
        month = feb,
          eid = {arXiv:1802.03426},
        pages = {arXiv:1802.03426},
archivePrefix = {arXiv},
       eprint = {1802.03426},
 primaryClass = {stat.ML}
}

@ARTICLE{1981MNRAS.194..809L,
       author = {{Larson}, Richard B.},
        title = "{Turbulence and star formation in molecular clouds}",
      journal = {\mnras},
         year = 1981,
        month = apr,
       volume = {194},
       number = {4},
        pages = {809--826},
          doi = {10.1093/mnras/194.4.809}
}

@ARTICLE{1987ARA&A..25...23S,
       author = {{Shu}, Frank H. and {Adams}, Fred C. and {Lizano}, Susana},
        title = "{Star formation in molecular clouds: observation and theory.}",
      journal = {\araa},
         year = 1987,
        month = sep,
       volume = {25},
        pages = {23--81},
          doi = {10.1146/annurev.aa.25.090187.000323},
       adsurl = {https://ui.adsabs.harvard.edu/abs/1987ARA&A..25...23S}
}

@ARTICLE{2004ARA&A..42..211E,
       author = {{Elmegreen}, Bruce G. and {Scalo}, John},
        title = "{Interstellar Turbulence I: Observations and Processes}",
      journal = {\araa},
         year = 2004,
        month = sep,
       volume = {42},
        pages = {211--273},
          doi = {10.1146/annurev.astro.41.011802.094859},
archivePrefix = {arXiv},
       eprint = {astro-ph/0404451}
}

@ARTICLE{2026ApJ...997..345Z,
       author = {{Zhao}, Mengke and {Li}, Guang-Xing and {Xu}, Duo and {Qiu}, Keping},
        title = "{Equation versus AI: Predicting Density and Measuring Width of Molecular Clouds by Multiscale Decomposition}",
      journal = {\apj},
         year = 2026,
        month = feb,
       volume = {997},
       number = {2},
          eid = {345},
        pages = {345},
          doi = {10.3847/1538-4357/ae3377},
archivePrefix = {arXiv},
       eprint = {2508.01130},
 primaryClass = {astro-ph.GA},
       adsurl = {https://ui.adsabs.harvard.edu/abs/2026ApJ...997..345Z}
}

@MISC{2022LeCunAMI,
       author = {{LeCun}, Yann},
        title = "{A Path Towards Autonomous Machine Intelligence}",
         year = 2022,
        month = jun,
          url = {https://openreview.net/forum?id=BZ5a1r-kVsf},
         note = {OpenReview, version 0.9.2}
}

@ARTICLE{2003MNRAS.345..325C,
       author = {{Cho}, Jungyeon and {Lazarian}, A.},
        title = "{Compressible Magnetohydrodynamic Turbulence: Mode Coupling, Scaling Relations, Anisotropy, Viscosity-damped Regime and Astrophysical Implications}",
      journal = {\mnras},
         year = 2003,
        month = oct,
       volume = {345},
       number = {1},
        pages = {325--339},
          doi = {10.1046/j.1365-8711.2003.06941.x},
archivePrefix = {arXiv},
       eprint = {astro-ph/0301062}
}

@ARTICLE{2011MNRAS.411.2067L,
       author = {{Li}, Hua-bai and {Blundell}, Raymond and {Hedden}, Abigail and {Kawamura}, Jonathan and {Paine}, Scott and {Tong}, Edward},
        title = "{Evidence for Dynamically Important Magnetic Fields in Molecular Clouds}",
      journal = {\mnras},
         year = 2011,
        month = mar,
       volume = {411},
       number = {3},
        pages = {2067--2075},
          doi = {10.1111/j.1365-2966.2010.17839.x},
archivePrefix = {arXiv},
       eprint = {1007.3312}
}

@misc{zhao2026scalevectoralignmentscaleawareframework,
      title={Scale-Vector Alignment: A Scale-Aware Framework for Spatially Resolved Morphological Similarity in Astronomical Images}, 
      author={Mengke Zhao and Guang-Xing Li and Keping Qiu and Shanghuo Li},
      year={2026},
      eprint={2609.24304},
      archivePrefix={arXiv},
      primaryClass={astro-ph.IM},
      url={https://arxiv.org/abs/2609.24304}, 
}

@misc{zhao2026gravitydrivenemergencemultifractaldensity,
      title={Gravity-driven Emergence of Multi-fractal Density Structure in the Orion A Integral Shaped Filament}, 
      author={Mengke Zhao and Guang-Xing Li and Keping Qiu and Guangya Zeng},
      year={2026},
      eprint={2609.24319},
      archivePrefix={arXiv},
      primaryClass={astro-ph.GA},
      url={https://arxiv.org/abs/2609.24319}, 
}

\appendix
\restartappendixnumbering

\section{MHD Simulation and Numerical Data}\label{app:simulation}

The analysis uses the self-gravitating, isothermal MHD turbulence simulations distributed through the Catalogue for Astrophysical Turbulence Simulations (CATS) \citep{2020ApJ...905...14B}. The calculations were performed with the constrained-transport MHD implementation in Enzo \citep{2010ApJS..186..308C} and were developed to study self-gravity and magnetic fields in supersonic molecular-cloud turbulence \citep{2012ApJ...750...13C,2015ApJ...808...48B}. We use the $256^3$ cube with initial sonic Mach number $M_{s,0}=9$ and initial plasma beta $\beta_0\simeq0.2$, at $t=0.9\,t_{\rm ff}$. This is the same simulation family and snapshot choice analyzed in the magnetic/kinetic-regime study of \citet{2025MNRAS.542.3246L}. The spatial sampling is $\Delta x=0.018$ pc, giving a box size of approximately 4.6 pc. These parameters are documented for the same data set in the published simulation analyses.

The main 2D analyses use the central $z=128$ slice. Figure~\ref{fig:framework}a shows the density field in number-density units; for the density baseline in Figure~\ref{fig:spectra} we convert mass density using $n=\rho/(\mu m_{\rm H})$ with $\mu=2.37$. The five-level CDD representation supplied to ScaleAware-JEPA is defined in Section~\ref{sec:sajepa}. The full three-dimensional cube is retained for the $8^3$-cell diagnostics in Figure~\ref{fig:manifold}.

\section{Analysis Definitions and Selection Rules}\label{app:estimators}

This appendix gives the analysis definitions used to connect the latent atlas to the physical diagnostics.

\paragraph{Latent anchors and spatial back-mapping.} Following the ScaleAware-JEPA latent-atlas procedure, representative positions are selected from localized regions of the three-dimensional UMAP projection and mapped back to their pixel-registered spatial coordinates \citep{2026arXiv260629723L}. The three anchors used here correspond to $(x,y)=(144,148)$, $(188,72)$, and $(128,78)$; their back-mapped density environments are described as clump-like, filament-like, and diffuse-like, respectively (Figure~\ref{fig:framework}a). The UMAP projection is used for interactive navigation, whereas the similarity back-maps are evaluated from cosine similarity in the full 32-dimensional latent representation. Figure~\ref{fig:spectra} uses the upper 33.33\% of each associated similarity field independently, while Figure~\ref{fig:mhdcal} uses an exclusive assignment to the strongest similarity above the corresponding 66.67th-percentile threshold.

\paragraph{Turbulent scaling.} For each Figure~\ref{fig:spectra} selection, $5\times10^6$ random pairs are drawn with replacement from the selected pixels. Pair separations use the minimum-image convention of the periodic domain and are binned in one-pixel intervals from 1 to 49 pixels. Equation~\eqref{eq:s2} is evaluated with pair weight $w_{ij}=\sqrt{\rho_i\rho_j}$, and the plotted quantity is $rS_{2,\rho}(r)$ against inverse separation $r^{-1}$. Power-law slopes are least-squares fits in log--log space over $5$--$25$ pixels.

\paragraph{Density-detrended state.} The local $\sigma_v$ and $M_A$ fields in Figure~\ref{fig:mhdcal} are evaluated in $8\times8$-pixel apertures using density-weighted velocity moments and periodic wrapping. The median $B(\rho)$ and $\sigma_v(\rho)$ trends are estimated in 24 equally spaced bins of $\log\rho$ between the 1st and 99th percentiles; a bin is retained if it contains at least 40 points. For the pairwise metrics, the three exclusive populations are downsampled to the same size. $D_M$ uses the mean difference and the average of the two covariance matrices. OVL is calculated from normalized 2D histograms over $-1\le\Delta\log B,\Delta\log\sigma_v\le1$, after Gaussian smoothing of the histogram grid. The HDR contours and one-dimensional $M_A$ curves are KDE visualizations and are not used to define the quoted medians.

\begin{figure}[ht!]
\centering
\includegraphics[width=\textwidth]{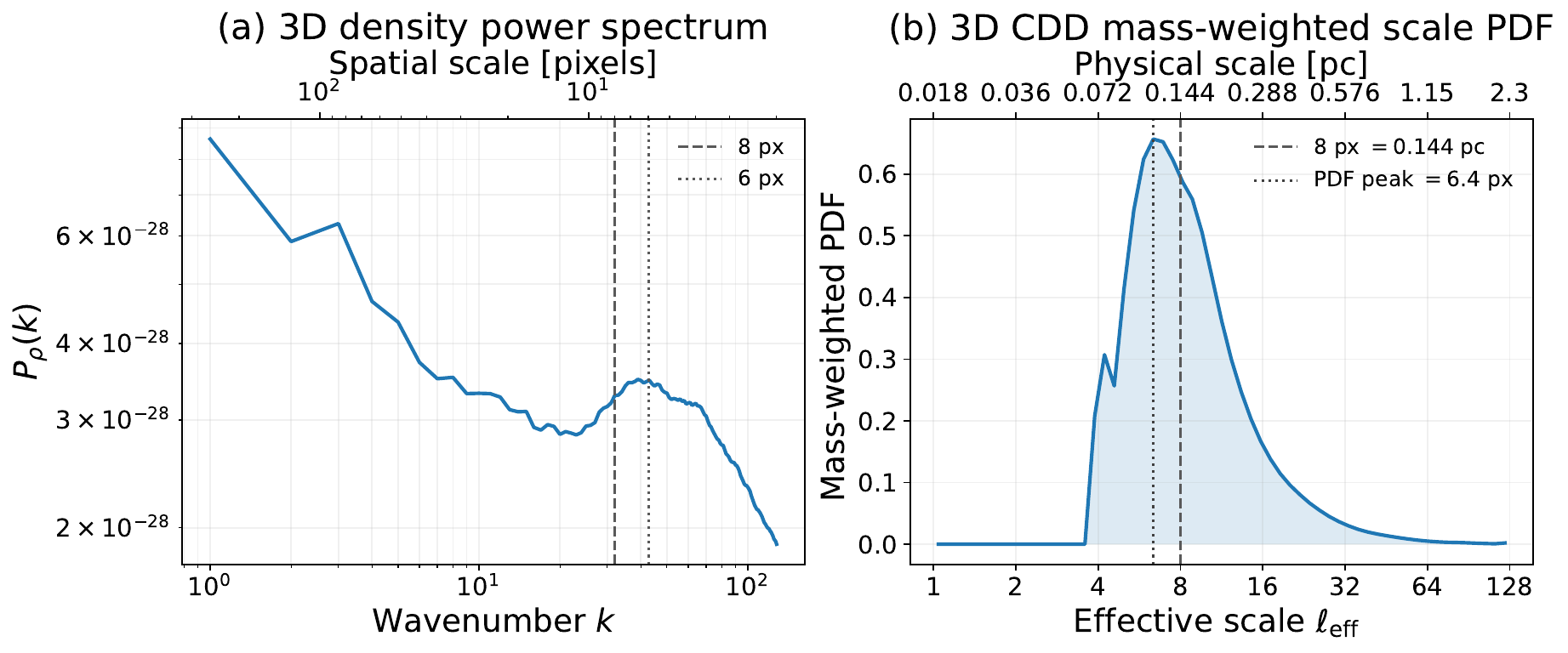}
\caption{Independent density-scale diagnostics from the full three-dimensional density cube. (a) Isotropic density power spectrum $P_\rho(k)$, with the small-scale spectral feature near $\sim6$ pixels. (b) Mass-weighted PDF of the three-dimensional CDD effective scale defined by Equation~\eqref{eq:seff}; the distribution peaks at $s_{\rm eff}\simeq6.4$ pixels. The dashed marker shows the adopted 8-pixel analysis scale, the nearest scale in the SA-JEPA CDD pyramid, corresponding to $\ell_*=0.144$ pc.}
\label{fig:scale_diag}
\end{figure}

\paragraph{Characteristic density scale and adopted analysis window.} The density field independently identifies the physical scale used for Figure~\ref{fig:manifold}. The three-dimensional density power spectrum, evaluated with the estimator of \citet{2024ApJ...976..209Z}, shows a local spectral maximum near $6$ pixels. We also apply CDD to the full three-dimensional density cube in logarithmic mode and denote the resulting scale-component amplitudes by $C_s(\bm{x})$. Following the logarithmic scale weighting of the volume-density mapping approach \citep{2026ApJ...997..345Z}, we define
\begin{equation}
 \log_2 s_{\rm eff}(\bm{x})=
 \frac{\sum_s C_s(\bm{x})\log_2 s}
      {\sum_s C_s(\bm{x})},
 \label{eq:seff}
\end{equation}
and construct its global PDF with the true volume density as the voxel mass weight. The distribution peaks at $6.4$ pixels (Figure~\ref{fig:scale_diag}). Both diagnostics therefore place the characteristic density scale near $6$--$6.4$ pixels. The SA-JEPA CDD pyramid uses $s=2,4,8,16,$ and $32$ pixels; we adopt its nearest scale, $s_*=8$ pixels, giving $\ell_*=8\Delta x=0.144$ pc.

\paragraph{Full-atlas balances and display.} The physical fields in Figure~\ref{fig:manifold} are evaluated in cubic $8^3$-cell neighborhoods using a uniform filter with periodic wrapping. Local turbulent velocity moments are density weighted,
\begin{equation}
 \bar\rho=\langle\rho\rangle,\qquad
 \bar v_q=\frac{\langle\rho v_q\rangle}{\bar\rho},\qquad
 \sigma_{v,3D}^2=\sum_{q=x,y,z}\left(\frac{\langle\rho v_q^2\rangle}{\bar\rho}-\bar v_q^2\right),
\end{equation}
while $\overline{\bm B}=\langle\bm B\rangle$ is the local mean magnetic field. For the virial and flux normalizations, the cube is represented by an equal-volume sphere with $R_{\rm eff}=(3/4\pi)^{1/3}\ell_*=0.089$ pc, $M=(4\pi/3)R_{\rm eff}^3\bar\rho$, $\Sigma=M/(\pi R_{\rm eff}^2)$, and $\sigma_{1D}^2=\sigma_{v,3D}^2/3$. Figure~\ref{fig:manifold} uses the saved 3D UMAP coordinates only for visualization. Neighborhoods are instead defined in the original 32-dimensional predicted latent vectors: after $L_2$ normalization, a cosine-neighborhood search is implemented as Euclidean nearest-neighbor search on the normalized vectors. A random subset of at most 18,000 valid points is displayed, and each point is colored by the median physical quantity among its 96 nearest latent neighbors. The global medians, 16--84 percentile ranges, and regime fractions reported in the text are calculated from the unsmoothed physical fields.

\end{document}